\documentclass[10pt, conference, letterpaper]{IEEEtran}

\usepackage{amsthm}
\usepackage{amssymb}
\usepackage{amsmath}
\usepackage{graphicx}
\graphicspath{{figs/}}
\usepackage[font=small]{caption}
\usepackage[labelformat=simple,font=scriptsize]{subcaption}

\usepackage[dvipsnames]{xcolor}
\usepackage{cite}
\usepackage{url}
\usepackage{algorithm}
\usepackage{algpseudocode}
\usepackage{comment}
\usepackage{relsize}
\usepackage{enumitem}
\usepackage{tikz}

\newcommand{\eg}{\textit{e.g.},}
\newcommand{\ie}{\textit{i.e.},}

\newcommand{\eqend}{\,.}

\newcommand{\cat}[1]{\medskip\noindent\textbf{#1.}}
\newcommand{\scat}[1]{\vspace{0.5ex}\noindent\textbf{#1.}}
\newcommand{\fcat}[1]{\noindent\textbf{#1.}}
\newcommand{\sys}{\textsmaller{\textsf{\mbox{Our Work}}}}
\newcommand{\bull}[1]{\vspace{0.5ex}\noindent\textbf{$\bullet$~#1:}}

\newcommand{\pr}[1]{\mathbb{P}\left\{#1\right\}}

\newcommand{\mx}[1]{\ensuremath{\mathbf{#1}}}
\newcommand{\mc}[1]{\mathcal{#1}}
\newcommand{\set}[1]{\left\{ #1 \right\}}

\newcommand*\circled[1]{\tikz[baseline=(char.base)]{%
            \node[shape=circle,fill=gray!20,draw,inner sep=1pt] (char) {#1};}}

\everydisplay{\small}

\begin{document}

\bstctlcite{IEEEexample:BSTcontrol}


\title{Shift Detection and Adaptation for Network Intrusion Detection}

\author{\IEEEauthorblockN{Ehssan Mousavipour}
	\IEEEauthorblockA{University of Calgary\\
		Calgary, Canada\\
		Email: ehssan.mousavipour@ucalgary.ca}
	\and
	\IEEEauthorblockN{Andrey Dimanchev}
	\IEEEauthorblockA{University of Calgary\\
		Calgary, Canada\\
		Email: andrey.dimanchev@ucalgary.ca}
	\and
	\IEEEauthorblockN{Majid Ghaderi}
	\IEEEauthorblockA{University of Calgary\\
		Calgary, Canada\\
		Email: mghaderi@ucalgary.ca}}
\maketitle

\begin{abstract}
Distribution shift, a change in the statistical properties of data over time, poses a critical challenge for deep learning anomaly detection systems. Existing anomaly detection systems often struggle to adapt to these shifts. Specifically, systems based on supervised learning require costly manual labeling, while those based on unsupervised learning rely on clean data, which is difficult to obtain, for shift adaptation. 
Both of these requirements are challenging to meet in practice.	
In this paper, we introduce \sys, a framework for supervised anomaly detection in network data that continually detects and adapts to distribution shifts in an online manner. \sys\ eliminates manual intervention through a novel pseudo-labeling technique and uses a knowledge distillation-based adaptation strategy to prevent catastrophic forgetting. Evaluated on three long-term network datasets, \sys\ demonstrates superior adaptation performance compared to state-of-the-art methods that rely on manual labeling, achieving F1-score improvements of up to $11.72\%$. This proves its robustness and effectiveness in dynamic networks that experience distribution shifts over time.
\end{abstract}

\section{Introduction}
\label{s:intro}

Network intrusion detection systems (NIDS) play a crucial role in detecting attacks and intrusions on a network or hosts within it~\cite{chandola2009anomaly}. Historically, NIDS primarily relied on signature-based intrusion detection, which identifies anomalies through comparing network data against databases of known attack patterns~\cite{garcia2009anomaly}. While effective against documented threats, this approach falls short when encountering unknown anomalies, also known as zero-day attacks. To overcome this limitation, deep learning (DL) models have been increasingly adopted for network anomaly detection~\cite{chalapathy2019deep, han2023anomaly, zhang2024aoc, zhang2025continual,mirsky2018kitsune, tang2020zerowall}. 
These models can be trained using supervised learning, which leverages labeled datasets consisting of normal data and known attacks~\cite{jordaney2017transcend, yang2021cade, andresini2021insomnia, barbero2022transcending}, or unsupervised learning, which identifies anomalies by learning the inherent patterns of normal data, also referred to as zero-positive learning~\cite{han2023anomaly}. The shift towards DL-based anomaly detection has led to significant advancements, establishing a more robust network security framework, particularly for detecting unknown types of attacks~\cite{garcia2009anomaly}.

Nevertheless, a common limitation with current DL-based anomaly detection solutions is the presumption of a static data distribution between training and deployment environments~\cite{han2023anomaly,zhang2024aoc, zhang2025continual}. Assuming the environments do not change diverges significantly from real-world network and system dynamics, where the interactions of users and systems evolve over time~\cite{sommer2010outside}. For instance, factors such as the introduction of new network protocols, system updates, or the patching of vulnerabilities can lead to substantial shifts in data distribution between the training and deployment environment~\cite{han2023anomaly,zhang2024aoc, zhang2025continual}. Consequently, solutions that do not account for such distributional drift exhibit degraded performance, resulting in an increase of both False Positives (FP) and False Negatives (FN) for DL-based anomaly detection systems. 

Addressing distribution shift, which is defined as a change in the statistical properties of testing data compared to the original training data, can be achieved through two strategies~\cite{zhang2025continual}: either continuously retraining the DL model with both new and old data~\cite{cretu2008casting, cretu2009adaptive, kantchelian2013approaches, pendlebury2019tesseract, jan2020throwing, nigenda2022amazon, zhang2024aoc}, or detecting and adapting the model to the shift~\cite{jordaney2017transcend, yang2021cade, andresini2021insomnia, barbero2022transcending, zhang2025continual}. In the context of cybersecurity, however, continuous model retraining is often impractical due to the extensive manual labeling required~\cite{pendlebury2019tesseract,jan2020throwing}. For supervised models, all new training data necessitates labeling for learning. Similarly, unsupervised models rely only on clean normal data for training, to establish a precise definition of “normal” behavior. Thus, the retraining strategy is inherently labor-intensive, costly, and also introduces the complex question of optimal model update frequency~\cite{xu2020method,meng2021logclass,yang2021cade,han2023anomaly}. Frequent updates impose significant resource demands, whereas delayed updates lead to performance degradation. In critical security applications, both are unacceptable~\cite{han2023anomaly}.

Consequently, the most practical approach in the security domain is to detect and adapt to the normality shift, defined as a change in the underlying data distribution of normal behavior for a system~\cite{han2023anomaly}, while preserving the model’s previously acquired knowledge to prevent catastrophic forgetting. Catastrophic forgetting occurs when a model, upon learning new information, experiences a significant degradation or loss of its prior knowledge~\cite{kirkpatrick2017overcoming, du2019lifelong, han2023anomaly}. 
For this approach, however, the challenge of manual labeling remains, as the adaptation step, whether for supervised or unsupervised methodologies, still involves a costly and time-consuming process of labeling new data~\cite{jordaney2017transcend,du2019lifelong,yang2021cade,barbero2022transcending,han2023anomaly}. Moreover, specialized techniques need to be applied to prevent catastrophic forgetting. 

To address the catastrophic forgetting challenge, some prior works~\cite{du2019lifelong, han2023anomaly} attempt to employ prevention strategies based on weight consolidation~\cite{kirkpatrick2017overcoming}. These methods estimate the importance of model parameters with respect to old knowledge and then use these weight assignments to penalize updates to critical parameters during adaptation. Nevertheless, such parameter importance estimation techniques are prone to exploding gradients, a phenomenon where gradients 
become excessively large during model updates. This well-known issue destabilizes the learning process and impedes effective adaptation~\cite{kutalev2021stabilizing, jastrzebski2021catastrophic}.
Alternatively, other works have explored a teacher-student paradigm, where a teacher model guides the student’s training during adaptation~\cite{li2017learninglwf, zhang2025continual}. This approach encourages the student to imitate the teacher’s output to overcome catastrophic forgetting. However, most implementations focus only on model outputs and overlook underlying structural relationships, offering an incomplete solution for substantial distribution shifts~\cite{li2017learninglwf, vahedifar2025no}, which could happen in real-world deployments~\cite{han2023anomaly}.

The preceding challenges lead to two central research questions in DL-based anomaly detection for NIDS:
\begin{enumerate}[itemindent=0pt, leftmargin=*, label=\small\protect\circled{\color{black}\arabic*}]
\item \textbf{Manual Labeling:} How to eliminate manual labeling in order to overcome the costly overhead of labeling?
\item \textbf{Catastrophic Forgetting:} How to overcome catastrophic forgetting effectively when adapting to new data?
\end{enumerate}

To address these questions, we introduce \sys, a supervised learning framework that is designed for anomaly detection in real-world scenarios based on detecting, explaining, and adapting to distribution shifts. \sys\ specifically aims to overcome manual labeling overhead and catastrophic forgetting.
To tackle the first challenge, we design a \textit{distribution-level voting mechanism} for assigning pseudo-labels to incoming unlabeled network data. Pseudo-labels are predicted labels generated by a model or a detection module trained on labeled data, which are then treated as ground-truth labels for downstream tasks. Our design employs an Autoencoder (AE) to learn data representations from both its encoder and decoder. We then model the similarity scores between these representations and a learned average normal representation to generate statistical distributions. This process yields distinct distributions for the normal and abnormal data, forming the basis of our voting system. Subsequently, we leverage Kullback-Leibler (KL) divergence to measure the similarity of the target distribution (abnormal) to the reference distribution (normal). This allows us to assess whether the encoder or decoder better distinguishes between normal and abnormal distributions and choose pseudo-labels accordingly. Unlike previous work that utilizes sample-wise mechanisms for pseudo-label assignment~\cite{zhang2024aoc}, our approach leverages the statistical properties of the learned data distributions, leading to more robust labels. 

To address the second challenge, we introduce \textit{a novel knowledge distillation loss function} that is integrated into the teacher-student framework. This specialized loss function enables a teacher model, trained on previous data, to distill its knowledge to a student model. This allows the student to more effectively integrate new data with prior knowledge, thereby mitigating catastrophic forgetting. The core principle of our loss function is to encourage the student model to preserve the teacher’s structure of pairwise similarities: sample pairs that produced similar representations in the teacher model are constrained to produce similarly structured representations in the student model. By minimizing the KL-divergence between these two similarity distributions, our approach ensures that crucial relational information is retained and integrated, thereby effectively mitigating the risk of catastrophic forgetting as the model undergoes updates. 

We conducted a comprehensive experimental evaluation using three prominent datasets in the field of NIDS: Kyoto2006+~\cite{song2011statisticalkyoto, druagoi2022anoshift}, CICIDS2017~\cite{cicids2017, sharafaldin2018toward}, and CICDDoS2019~\cite{cicddos2019, DDoS}. We benchmarked \sys\ against state-of-the-art continual learning anomaly detection frameworks such as OWAD~\cite{han2023anomaly}, a framework based on weight consolidation, and Strategic Selection and Forgetting (SSF)~\cite{zhang2025continual}, which uses a memory replay strategy. For our pseudo-labeling component, we compared our novel distribution-level voting mechanism against the sample-wise approach employed by AOC-IDS~\cite{zhang2024aoc}. Our experimental results demonstrate that Our Work significantly outperforms both OWAD (by up to 11.72\%) and SSF (by up to 7.42\%) in terms of F1-score.

Our contributions in this work are summarized as follows:
\begin{itemize}
	\item We present the design of \sys, a deep anomaly detection system for dynamic network environments, capable of online shift detection and adaptation.
	\item We propose a novel distribution-level voting mechanism to generate robust pseudo-labels for shift adaptation in \sys, eliminating the need for manual labeling. 
	\item We design a novel knowledge distillation loss function to mitigate catastrophic forgetting in \sys\ by preserving the relational structure between data samples.
	\item We evaluate \sys\ via extensive empirical validation on three NIDS datasets, demonstrating that \sys\ significantly outperforms state-of-the-art solutions.
\end{itemize}

The rest of the paper is organized as follows. Section~\ref{s:design} presents the \sys's design. Evaluation results are presented and discussed in Section~\ref{s:eval}. Section~\ref{s:related} reviews the related work, while Section~\ref{s:conc} concludes the paper.
\section{Design}
\label{s:design}

In this section, we detail the components of our design. The workflow begins when data arrives at the Anomaly Detection Module (ADM) for pseudo-labeling. This is followed by a three-step process to address distribution shift: Shift Detection, Shift Explanation, and Model Adaptation.

\subsection{Anomaly Detection Module}

In a real-world environment, benign network traffic follows more consistent patterns than malicious traffic, which can vary significantly. We leverage this property using contrastive learning, a method designed to pull representations of similar samples (\ie\ normal traffic) together while pushing dissimilar samples (\ie\ malicious traffic) farther apart in the embedding space~\cite{tian2020makescontrastive, khosla2020supervisedcontrastive, zhang2024aoc}. The outcome is a coherent and stable representation of normal samples in the embedding space, ideal for use as discriminative anchors to push away representation of abnormal samples from the normal samples representation. Thus, our ADM consists of two stages: 1) representation learning using contrastive learning, and 2) anomaly detection based on pseudo-label generation for new samples.

\cat{Representation Learning}
To detect anomalies, we first need to establish a comprehensive representation of normal data. To achieve this, we leverage InfoNCE~\cite{oord2018representation, hoffmann2022ranking} loss function to generate a prototypical representation of normal data.
Hereafter, we use the following notations: the subscripts `$n$' and `$a$' denote normal and abnormal data, while `$en$' and `$de$' denote the encoder and decoder of our Autoencoder (AE), respectively. 
Let $\mc{X}_n (\mc{X}_a)$ and $\mc{Y}_n (\mc{Y}_a)$ denote the set of normal (abnormal) input data samples and their labels, respectively. The dataset contains $l_n$ normal and $l_a$ abnormal samples. Each sample $\mx{x}$ has $d \in \mathbb{N}$
features and is labeled by $y \in \{0, 1\}$, where $0$ represents normal instances and $1$ represents abnormal instances in our dataset. We refer to the learned representation for a single sample from the autoencoder model $\mathrm{AE}_{\theta}$ (parameterized by $\theta$) as an embedding, denoted by $\mx{e}$. The set of all such embeddings is denoted by $\mc{E}$, where $\mx{e} \in \mc{E}$.


Our goal is to pull representations of normal samples, also known as positive pairs $\{\mx{e}_{n,i},\mx{e}_{n,j}\}$, for $i,j  \in \{1,...,l_n\}$,
closer together in the embedding space while pushing representations of abnormal samples, known as negative set
\begin{math}
	\{\mx{e}_{a,k} | k \in \{1,2,...,l_a\}\}
\end{math},
away from those of the normal samples. To achieve this, we employ the InfoNCE loss function, where we treat the representation of every normal sample $\mx{e}_{n,i}$
as an anchor. For each anchor, every other normal sample $\mx{e}_{n,j}$
($j \ne i$) is a positive counterpart, while all abnormal samples serve as the negative set
for that anchor. Thus, the loss function $\mc{L}_{i,j}$ is given by~\cite{oord2018representation, hoffmann2022ranking}:
\begin{equation}
	\mc{L}_{i,j} = -\log\ {\frac{\exp(h(i,j)/\tau)}{\exp(h(i,j)/\tau) \ + \ \sum_{k=1}^{l_a}\exp(h(i,k)/\tau)}}, \label{eq:contrast_loss}
\end{equation}
where $h(i,j)$ is the function for computing similarity score and $\tau$ is the temperature value, which controls how strictly the model must distinguish between similar and dissimilar samples, and accepts values between 0 and 1. The similarity function used in our design is cosine similarity~\cite{schutze2008introduction} defined as $h(i,j) = \frac{\mx{e}_{i} \cdot \mx{e}_{j} }{\|\mx{e}_i\| \|\mx{e}_j\|}$. The average loss $\mc{L_C}$ is then calculated by averaging the individual loss terms $\mc{L}_{i,j}$ over all possible normal anchor-positive pairs, as follows:
\begin{equation}
	\mc{L_C} = \frac{1}{l_n(l_n \ - 1)}
				\textstyle\sum_{i=1}^{l_n}\ 
				\textstyle\sum_{j=1, \ j\ne i}^{l_n} \mc{L}_{i,j} 
				\label{eq:pair_loss}
	\eqend
\end{equation}
%
The total loss is defined as the sum of the encoder and decoder average losses:
\begin{equation}
	\mc{L}_{\text{total}} = \mc{L}_C^{en} \ + \ \mc{L}_C^{de}, 
	\label{eq:final_loss}
\end{equation}
where $\mc{L}_C^{en}$ and $\mc{L}_C^{de}$ are calculated using~\eqref{eq:pair_loss} for the encoder and decoder, respectively.
The loss $\mc{L}_{\text{total}}$ helps the model $\mathrm{AE}_\theta$ to learn embeddings that increase similarity for normal-normal pairs, while decreasing it for normal-abnormal pairs.

\cat{Pseudo Label Generation}
%
%
Once the model is trained, a prototypical representation of the positive class can be established, typically by computing the mean of all sample representations from that class, denoted by $\mx{\bar{e}}_{pos} $. Subsequently, the similarity score of every training sample to this prototype is calculated. This yields two distinct sets of scores, one for samples belonging to the positive class and one for samples belonging to the negative class. We model these two sets of scores using distinct Gaussian distributions. This is a well-established technique, as Gaussian Mixture Models are foundational for modeling data distributions~\cite{Gaussian, mirsky2018kitsune, zhang2024aoc}. 
To find the optimal parameters for the two distributions, we employ Maximum Likelihood Estimation (MLE)~\cite{seydgar2022semisupervised, zhang2024aoc}. This process is performed independently for the representations from both the encoder and decoder, yielding two sets of parameters, namely $(\mu_{n}^{en}, \sigma_{n}^{en})$ and $(\mu_{a}^{en}, \sigma_{a}^{en})$ for the encoder, and $(\mu_{n}^{de}, \sigma_{n}^{de})$ and $(\mu_{a}^{de}, \sigma_{a}^{de})$ for the decoder. This produces two pseudo-labels for each sample, one by the encoder and one by the decoder. To choose one of the labels, one approach is to compare the absolute difference between likelihoods for every single sample, as employed in~\cite{zhang2024aoc}. However, at a distribution level, this sample-wise method cannot determine whether the encoder or decoder better separates normal from abnormal samples.




To this end, we introduce a novel distribution-level voting mechanism. 
Our key intuition is that greater separability indicates more confidence, and thus more reliable pseudo-labels.
We quantify separability using the Kullback-Leibler (KL) divergence between the two learned Gaussian distributions for each component (encoder or decoder), as presented below:
\begin{equation}
	\text{KL}^{mod} = \text{KL}\big(\mc{N}(\mu_{a}^{mod},\sigma_{a}^{mod}) \ \| \ \mc{N}(\mu_{n}^{mod},\sigma_{n}^{mod})\big),
\end{equation}
where $mod \in \set{en, de}$. For any given evaluation run, these KL scores are calculated once based on the distributions fitted to the training set, yielding two fixed scalar values for each component, \ie\ $\text{KL}^{en}$ and $\text{KL}^{de}$. These scores are therefore independent of any individual test sample.
The final classification of a test sample proceeds in two steps. First, for an unlabeled sample, the encoder and decoder generate initial labels $(\hat{y}_{en}, \hat{y}_{de})$ based on its similarity score, $s_{test}$, to the average normal representation:
\begin{equation}
	\hat{y}_{mod} = 
	\begin{cases}
		0, & \text{if } 
		\scalebox{0.85}{$\pr{s_{test} | \mc{N}(\mu_{n}^{mod}, \sigma_{n}^{mod})} > \pr{s_{test}| \mc{N}(\mu_{a}^{mod}, \sigma_{a}^{mod})}$}, \\
		1, & \text{otherwise},
	\end{cases}
\end{equation}
where $mod \in \set{en, de}$.
The final label \(\hat{y}\) is chosen from the component with higher, pre-calculated KL-divergence score:
\begin{equation}
	\hat{y} =
	\begin{cases}
		\hat{y}_{en}, & \text{if } \text{KL}^{en} > \text{KL}^{de}, \\ \label{eq:final_label}
		\hat{y}_{de}, & \text{otherwise} \eqend
	\end{cases}
\end{equation}
We refer to the distribution parameters from the selected component, \ie\ encoder or decoder, as $\mc{N}(\mu_n^{select}, \sigma_n^{select})$ and $\mc{N}(\mu_a^{select}, \sigma_a^{select})$. Our voting strategy offers a distribution-level perspective and provides a clear justification for selecting the final label, which we will demonstrate in our experiments.

\subsection{Shift Detection, Explanation, and Adaptation}
\fcat{Shift Detection}
A key step in this process is applying normality shift detection, a technique previously used in unsupervised learning~\cite{han2023anomaly}, to our supervised framework. For models relying on a discriminative boundary, this is non-trivial; our approach, however, is well-suited for this task. Even though, our model is trained with both normal and abnormal data, the primary objective of our contrastive learning process is to learn a representation of normality. Because our model fundamentally learns the characteristics of normal data, normality shift detection can still be applied in our supervised paradigm.

To achieve this, we first generate distributions from the similarity scores of old and new samples. We then use these distributions and Bayes' Theorem to calculate the posterior probability of a sample being normal, \ie\ $\pr{\text{normal} | s_i}$ (where $s_i$ is the similarity score of a sample's representation to the average normal representation):
\begin{equation}
	\pr{\text{normal} | s_i} =
	\frac{\pr{s_i | \text{normal}} \cdot \pr{\text{normal}}}{\pr{s_i}}, \label{eq:posterior_pr}
\end{equation}
where, $\pr{s_i | \text{normal}} = \pr{s_i | \mc{N}(\mu_{n}^{select}, \sigma_{n}^{select})}$, and,
\begin{align}
	\pr{s_i} &= \pr{s_i | \mc{N}(\mu_{n}^{select}, \sigma_{n}^{select})} \cdot \pr{\text{normal}} \nonumber\\
	 &\quad + \pr{s_i | \mc{N}(\mu_{a}^{select}, \sigma_{a}^{select})} \cdot \pr{\text{abnormal}} \label{eq:s_pr}
	 \eqend
\end{align}
The prior probabilities, $\pr{\text{normal}}$ and $\pr{\text{abnormal}}$, are derived from the class proportions from the training data.

With the posterior probabilities calculated for both the old and new samples, 
we perform a statistical comparison to detect a shift. We represent the two probability distributions using frequency histograms (over the same bins) 
and compare them using permutation tests. This non-parametric test is particularly effective as it makes no assumptions about the data's underlying distribution. We use the KL-divergence as the test statistic to evaluate the null hypothesis ($H_0$) of no distribution shift against the alternative ($H_1$) that the distributions differ.

\cat{Shift Explanation}
The goal of shift explanation is to identify a representative subset of samples from both the old and new distribution that best describes the shift. Our approach is built upon the optimization framework proposed in~\cite{han2023anomaly}. However, because our framework provides autonomous pseudo-labeling, we relax the constraint on minimizing manual labeling costs inherent in the original approach and replace it with a new constraint designed to minimize the total number of selected samples, with the practical goal of reducing the computational training cost during model adaptation.

To this end, we define two learnable mask vectors $\mx{m}^{o}$ and $\mx{m}^{n}$, corresponding to the old ($\mc{X}^{o}$) and new ($\mc{X}^{n}$) sample sets. If the sample ($\mx{x_i}$) is selected for explaining the shift, then the entry $\mx{m}_i$ in these vectors is set to $1$, and $0$, otherwise. 
The optimization objective is to minimize a weighted sum of three terms: an accuracy loss ($\mc{L}_{acc}$), a computation loss ($\mc{L}_{compute}$), and a determinism loss ($\mc{L}_{det}$), as given below:
\begin{equation}
	\min_{\mx{m}^{o} \oplus \mx{m}^{n}} \mc{L}_{acc} + \lambda_1 \mc{L}_{compute} + \lambda_2 \mc{L}_{det},\label{eq:opt}
\end{equation}
where $\lambda_1$ and $\lambda_2$ control the relative contribution of each loss.
Since solving integer optimization problems is generally NP-hard, we first relax the binary variables $\mx{m}_i$ by assuming $\mx{m}_i \in [0,1]$ and solve the problem. Then, we round the solution of the relaxed problem to obtain the binary mask vectors.

\bull{Accuracy Loss} This term's purpose is to ensure that the samples selected by the masks can accurately reconstruct the new distribution. Let $P^n$ and $P^o$ denote the distributions of posterior probabilities $\pr{\text{normal} | s_i}$, as given by~\eqref{eq:posterior_pr}, for the new and old samples, respectively, represented as histograms consistent with the shift detection phase. The accuracy loss is defined as the KL-divergence between the new and reconstructed distribution from the selected samples by the masks:
\begin{equation}
	\mc{L}_{acc} = \text{KL}
					\big(P^n \| (\mx{m}^o \odot P^o) \oplus (\mx{m}^n \odot P^n) \big), 
\end{equation}
where $\oplus$ and $\odot$ denote concatenation and element-wise product, respectively. Here, the term $(\mx{m} \odot P)$ is a shorthand notation for building a new histogram from the posterior probabilities of samples selected by the mask $\mx{m}$.

\bull{Computation Loss}
Our goal is to minimize the training cost by selecting the minimum number of samples from the old and new data. This contrasts with AOC-IDS~\cite{zhang2024aoc}, which incurs a high computational cost by retraining on all samples. Thus, we define the following loss function:
\begin{equation}
	\mc{L}_{compute} = \|\mx{m}^o \oplus \mx{m}^n\|_1,
\end{equation}
which would penalize the model for every sample it selects, forcing it to find the smallest set of samples for adaptation.

\bull{Determinism Loss} This term ensures that the selection process is confident (\ie\ not random). We utilize the binary entropy of the mask vectors:
\begin{equation}
	\mc{L}_{det} = \mathop{\mathbb{E}}_{m \in \mx{m}^o \oplus \mx{m}^n} 
	\big[-m \log (m) - (1 - m) \log (1-m)\big],
\end{equation}
which is minimized when the mask values are close to 0 or 1 (indicating a certain selection or rejection).

\cat{Shift Adaptation}
Having identified the minimal set of samples that explain the shift, the model must be updated accordingly. However, naively fine-tuning the model only on these selected samples would cause it to forget previously learned patterns, also known as catastrophic forgetting~\cite{kirkpatrick2017overcomingcontinual}.
%
%
To address this, we propose a novel adaptation strategy inspired by relational knowledge distillation technique~\cite{tung2019similarity}. We introduce a new knowledge distillation loss function that constrains the updated model to maintain the pairwise similarity structure of the pre-adaptation model. By adding this knowledge distillation loss to the original contrastive loss, we achieve a stable adaptation process that learns the new distribution while retaining the previously learned representations.

Our adaptation strategy employs a teacher-student framework where we first clone the trained model (teacher), which is then frozen. The copy serves as the student model, initialized with the teacher's weights, and subsequently fine-tuned.
The process for calculating our knowledge distillation loss is as follows: for each anchor sample $i$ within the $n$ selected samples that explain the shift, we calculate the pairwise similarity scores between the anchor $i$ and all other samples $j$ in the batch for both the teacher ($h_{teach}(i,j)$) and the student ($h_{stu}(i,j)$). Then, we convert these sets of similarity scores into probability distributions using the Softmax function. We denote the teacher's distribution over samples $j$ for a given anchor $i$ as $P_{teach}(\cdot | i)$. This distribution is defined as:
\begin{equation}
	P_{teach}(j|i) = \frac{\exp(h_{teach}(i,j))}{\sum_{k=1, k\neq i}^{n} \exp(h_{teach}(i,k))}
	\eqend
\end{equation}
The knowledge distillation loss \(\mc{L}_\text{KD}\), is the average KL-divergence between the student's and teacher's similarity distributions, calculated for each of the \(n\) anchor samples:
\begin{equation}
	\mc{L}_\text{KD} = \frac{1}{n} \textstyle\sum_{i=1}^n \text{KL}\big( P_{teach}(\cdot | i) \ \| \ P_{stu}(\cdot | i)\big)
	\eqend
\end{equation}
This \(\mc{L}_\text{KD}\) is computed for both the encoder and decoder representations of our AE, denoted by $\mc{L}_\text{KD}^{en}$ and $\mc{L}_\text{KD}^{de}$, respectively. The final loss for adaptation is a weighted sum of this distillation loss and the original contrastive loss defined in~\eqref{eq:final_loss} on the selected samples by the explanation phase:
\begin{equation}
	\mc{L}_\text{adapt} = (\mc{L}_{C}^{en} + \mc{L}_{C}^{de}) + \gamma \ (\mc{L}_\text{KD}^{en} + \mc{L}_\text{KD}^{de}), \label{eq:kl_adapt}
\end{equation}
where, the hyperparameter \(\gamma \in [0, 1]\) controls the contribution of the knowledge distillation loss.
This knowledge distillation loss guides the adaptation by ensuring the structural relationships learned by the teacher are preserved. In contrast to methods that penalize parameter changes~\cite{du2019lifelong, han2023anomaly}, which can lead to unstable gradients, our approach provides a more stable and effective adaptation process. Moreover, unlike frameworks that merely imitate raw outputs~\cite{zhang2025continual}, preserving the pairwise similarity structure ensures a more robust knowledge transfer.

\section{Evaluation}
\label{s:eval}

In this section, we present evaluation results for the post-adaptation performance of \sys\ against key baselines and include an ablation study of its core components.

\subsection{Setup and Design}
We use PyTorch and Python to implement \sys\ and its modules. The experiments were conducted on a desktop computer equipped with an NVIDIA GeForce RTX 3080 Ti graphics card, a 3.80~GHz Intel i7 CPU, and 16~GB of RAM.

\scat{Datasets and Preprocessing}
This work considers a more practical real-world scenario where the behavior of users and their interactions with a system evolve over a prolonged period of time, leading to the performance degradation of DL-based NIDS. A significant limitation in existing literature (\eg~\cite{du2019lifelong,zhang2024aoc,zhang2025continual}) is relying on short-term datasets, with time spans of merely hours to a few weeks. 
Such limited temporal scopes are insufficient to get an adequate assessment of distribution shifts over the long-term. To overcome this limitation, \sys\ is evaluated using datasets collected over extended periods of time. We leverage Anoshift~\cite{druagoi2022anoshift}, a known benchmark for NIDS application, to preprocess Kyoto2006+~\cite{song2011statisticalkyoto} and follow the data preparation methodology employed in~\cite{han2023anomaly}. 
All the datasets and benchmarks used in our evaluation are publicly available, facilitating reproducibility and enabling future research to build upon our findings. Additionally, during our literature review, we noticed the distributional shifts within the Kyoto2006+~\cite{song2011statisticalkyoto, druagoi2022anoshift} are not explained by the creators, leading us to use CICIDS2017~\cite{cicids2017, sharafaldin2018toward} and CICDDoS2019~\cite{cicddos2019, DDoS} to further validate \sys. The following subsections detail three public datasets, chosen to better evaluate \sys\ due to their documented and explainable distributional shifts.

\bull{\textbf{Kyoto2006+}}
For our first dataset, we utilized the Kyoto2006+~\cite{song2011statisticalkyoto} dataset, which was used in the Anoshift~\cite{druagoi2022anoshift} benchmark. This dataset consists of daily network traffic captured from both internal and external networks of Kyoto University, spanning approximately ten years from 2006 to 2015. Its feature set includes network flow information combined with feedback from security applications, such as antivirus and IDS. To ensure consistency with prior works, our preprocessing steps followed those employed by Anoshift. 

\bull{\textbf{CICIDS2017 and CICDDoS2019}}
To further assess the effectiveness of \sys, we incorporated the CICIDS2017~\cite{cicids2017, sharafaldin2018toward} and CICDDoS2019~\cite{cicddos2019, DDoS} datasets. These datasets enabled us to cross-validate our findings, evaluate performance on data collected over an extended period of time, and test with traffic more representative of real-world network conditions. The CICIDS2017 dataset contains both benign traffic, traffic analysis metrics, and a comprehensive set of contemporary network attacks, designed to mirror realistic network traffic. Similarly, CICDDoS2019 provides benign traffic enriched with network flow metrics, but its abnormal samples focus on common DDoS attacks. Both datasets feature 84 attributes and underwent a cleaning process to eliminate discrepancies and ensure structural consistency. Since the 2017 portion emphasizes TCP-based attacks, our preprocessing pipeline filtered out non-TCP traffic from CICDDoS2019 portion. 

During the preprocessing step, IP addresses were encoded using a byte-level method to transform them into a format compatible with machine learning models~\cite{IP}. 
A similar encoding was applied to the \texttt{flow\_ID} column. We selected this binary encoding due to its low computational overhead and its satisfactory performance in numerically representing IP addresses. Future work may explore other methods; our goal was to minimize training time while obtaining viable numeric representations, thereby maintaining focus on the models themselves. The timestamps were encoded using the \texttt{to\_datetime()} function from the pandas library~\cite{pandas_time_encoding}. Following preprocessing, the 2017 dataset exhibited a class distribution of 1,270,155 normal samples (69.53\%) and 556,549 abnormal samples (30.47\%). In contrast, the 2019 dataset showed a severe class imbalance, with only 26,745 normal samples (0.65\%) and 4,279,618 abnormal samples (99.35\%).

The pronounced class imbalance in the 2019 dataset would result in a biased testing set and thus would not produce a reliable measure of the model's performance, as it does not resemble the real-world scenario. Consequently, the 2019 dataset was synthetically balanced as an additional preprocessing step. For this purpose, both ADASYN~\cite{Chen_2021} and SMOTE~\cite{Chawla_2002}, two common methods of synthetically balancing classes, were tested. The synthetic generation was done in chunks. However, the class imbalance was often too extreme for ADASYN to work, so SMOTE was used as the fallback for those chunks. A second trial was conducted using SMOTE exclusively for synthetic data generation. After applying these methods, the 2019 dataset achieved a similar class distribution to the 2017 dataset, with 1,270,155 normal samples, of which 1,245,236 were synthetic and 24,919 were original. Following class balancing, we leveraged t-Distributed Stochastic Neighbor Embedding (T-SNE)~\cite{t-sne} visualizations to assess the correctness of the synthetic samples generated by SMOTE and ADASYN. 
These visualizations helped assess the structural and distributional fidelity of the synthetic samples.
Our analysis confirmed that both synthetic data generation methods accurately reproduced the features and underlying distribution of the original data. 
This allows using the synthetic samples for the evaluation step without compromising quality. 

\scat{Training and Testing Sets}
For the Kyoto2006+~\cite{song2011statisticalkyoto, druagoi2022anoshift} dataset, our data split closely followed the methodology employed by OWAD~\cite{han2023anomaly} to replicate their experimental setup. Specifically, the training portion from 2007 (designated as @T0) was utilized for model training. The training portion of the 2011 dataset was used for shift adaptation. We then evaluated \sys\ using the testing set of years 2011 (@T1), 2012 (@T2), 2013 (@T3), and 2014 (@T4). While we aimed to faithfully reproduce OWAD's data preparation, their publicly available code only included preprocessed 2007 and 2011 data, which had a different number of features compared to what the Anoshift benchmark preprocessing pipeline generates. Consequently, we tried to maintain record counts for 2007 and 2011 consistent with OWAD's code, while for the remaining portions, we relied on Anoshift's preprocessing steps. 

In a separate experimental setup focusing on the CIC datasets, the CICIDS2017~\cite{cicids2017, sharafaldin2018toward} dataset was designated for initial training, with CICDDoS2019~\cite{cicddos2019, DDoS} serving as the test set. In this test setup, @T1 refers to the model state before adaptation and @T2 represents the state after adaptation. To facilitate this, we partitioned the CICDDoS2019 data into an 80\% training split, used for shift detection and adaptation, and a 20\% testing split, used for the final evaluation at @T2. This particular train-test split emulates approaches seen in prior research while also having a two-year temporal gap, allowing us to assess the models' performance over a long period.

\scat{Baseline Methodologies}
We evaluated \sys\ against the following approaches: 

\bull{Retrain} Is a variation of \sys\ that retrains the model with a combined dataset of both previous and newly acquired data. Since the model is retrained, this method does not need to apply any shift detection or adaptation.

\bull{OWAD~\cite{han2023anomaly}} Is related to our work in terms of shift detection, explanation, and adaptation. However, they leverage zero-positive learning, an unsupervised machine learning technique. Their unsupervised approach presents the challenge of creating a direct performance comparison of their work with our supervised framework, due to the differing metrics used in supervised versus unsupervised learning. To overcome this misalignment, we use additional metrics, as detailed below.

\bull{AOC-IDS~\cite{zhang2024aoc}} Is a supervised method that continuously retrains itself at fixed intervals without going offline. This mimics real-world scenarios where an anomaly detection system might get updated periodically.

\bull{SSF~\cite{zhang2025continual}} Is another model that integrates shift detection, explanation, and adaptation by incorporating a buffer memory which is continually updated with representative samples. A key distinction of SSF, when compared to \sys, lies in its adaptation strategy: the model is updated regardless of whether a shift has occurred or not. In the absence of a shift, SSF adds a regularization term to the original loss function to prevent catastrophic forgetting. If a shift is detected, the model is updated without this regularization term. SSF incorporates a continual update mechanism, similar to AOC-IDS, but performs an additional model update when a shift is detected.

\scat{Experiment Settings}
The AE architecture, batch sizes, and shift explanation parameters follow OWAD~\cite{han2023anomaly}. We use the Adam optimizer for initial training (10 epochs, learning rate of 0.00001) and Stochastic Gradient Descent (SGD) for adaptation (5 epochs for Kyoto2006+, 10 for CIC; learning rate of 0.0001). For the contrastive loss functions used in~\eqref{eq:contrast_loss}, we set the temperature $\tau=0.02$. The distillation weight for the adaptation loss in~\eqref{eq:kl_adapt} is $\gamma = 0.1$. The hyperparameters used in~\eqref{eq:opt} are set to $\lambda_1=10$ and $\lambda_2=1$.

\scat{Metrics}
We primarily use \textit{Accuracy} and \textit{F1-score} to evaluate \sys's performance, as they effectively indicate the model's overall performance and its ability to distinguish between normal and abnormal samples. However, they are not suited for zero-positive learning methodologies, such as OWAD, which we use to compare \sys\ against. Because these models are exclusively trained on normal data, they require defining a threshold to classify new data as either normal or abnormal. Consequently, the most appropriate metrics for these methodologies are True Positive Rate (TPR) and False Positive Rate (FPR), which are used to construct an AUC-ROC curve by varying the detection threshold.

Since we use pseudo-labeling, and our model is supervised, it doesn't rely on a defined threshold to identify anomalies. Given that we cannot define thresholds from which we would construct an AUC curve, a direct comparison with OWAD's unsupervised model is difficult using these standard metrics. To overcome this, we have introduced two additional metrics to provide a more holistic and fair comparison across the different learning strategies. The first one is \textit{Balanced Accuracy (BACC)}~\cite{MCC}, calculated by taking the average of TPR and True Negative Rate (TNR). The motivation behind using this metric is to accurately evaluate \sys\ despite the class imbalances in our datasets. In such cases, a model might become biased towards the majority class, leading to a deceptively high Accuracy score. Conversely, BACC would drop in this scenario, indicating the model is overfitting. Since both TPR and TNR can be easily computed for supervised and unsupervised learning, BACC allows for a fair comparison. \textit{Matthews Correlation Coefficient (MCC)}~\cite{MATTHEWS1975442} is the second added metric, which is a statistical measure for binary classification that quantifies the correlation between a model's predictions and the ground-truth labels. Its score ranges from -1 to 1; where -1 signifies predictions are opposite to the ground-truth, 0 indicates random predictions, and 1 represents perfect alignment with the ground-truth labels. MCC can also be applied irrespective of the learning technique. For fairness, OWAD's performance is reported at its optimal threshold.

\subsection{Component Evaluation}
In this section, we present several experiments to demonstrate the effectiveness of the components designed for \sys\ compared to previously established methodologies.

\scat{Pseudo-Labeling} 
We study the effectiveness of distribution-level and point-wise voting mechanism for assigning pseudo labels in \sys\ and AOC-IDS, respectively.
As previously detailed, AOC-IDS decides to accept the pseudo labels from either the encoder or decoder based on how well either one is able to distinguish between normal and abnormal distributions for each sample. 
\sys, on the other hand, utilizes KL-divergence to assess which component (encoder or decoder) better separates normal and abnormal distributions across all samples. The component exhibiting higher class separation is selected for the pseudo-labeling. 

\begin{figure}[t]
	\centering
	\includegraphics[width=0.33\textwidth]{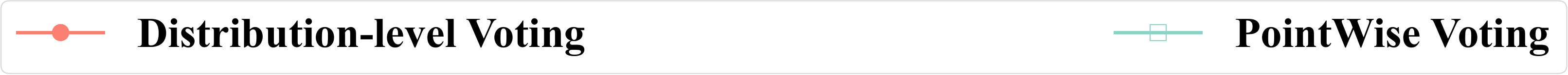}
	\begin{subfigure}[t]{\columnwidth}
		\centering
		\begin{subfigure}[t]{0.24\columnwidth}
			\centering
			\includegraphics[width=\textwidth]{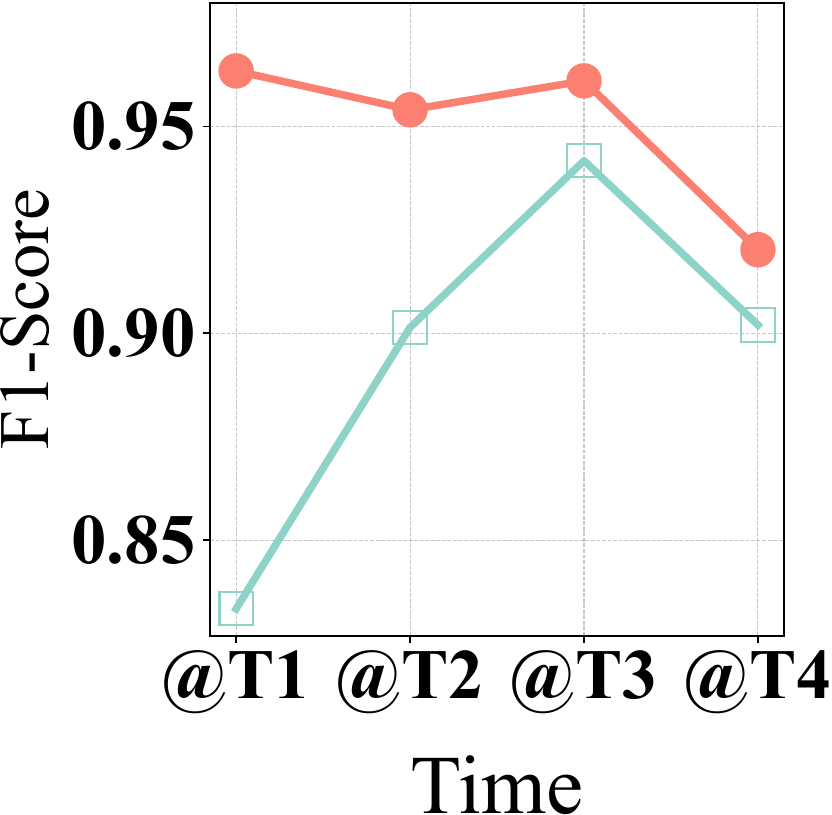}
			\caption{F1-Score}
			\label{f:smart:f1_kyoto_pointwise}
		\end{subfigure}
		\hfill
		\begin{subfigure}[t]{0.24\columnwidth}
			\centering
			\includegraphics[width=\textwidth]{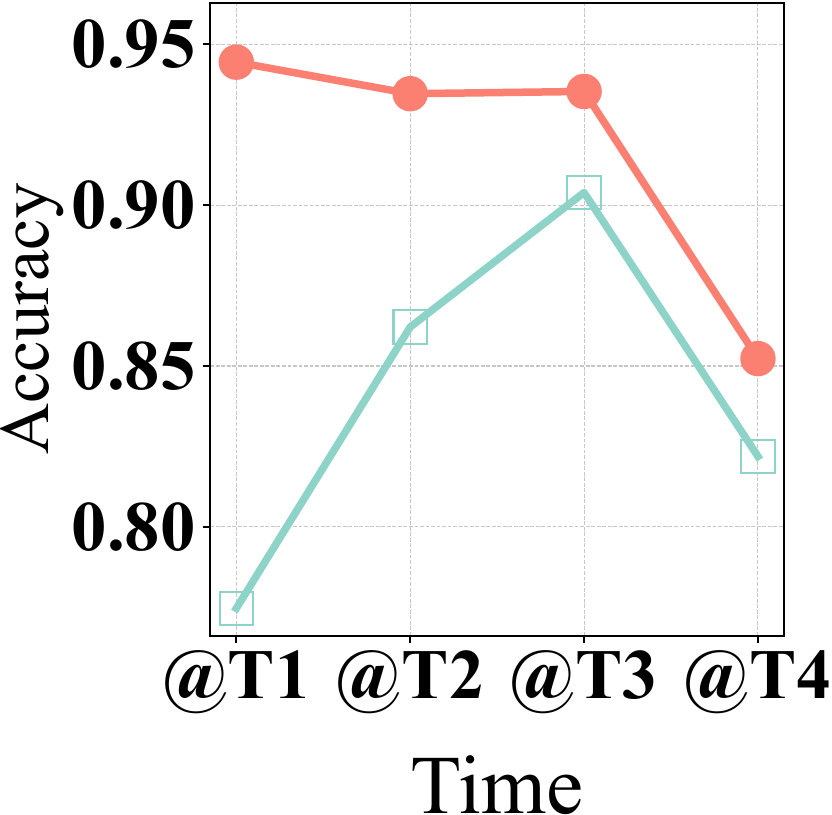}
			\caption{Accuracy}
			\label{f:smart:acc_kyoto_pointwise}
		\end{subfigure}
		\hfill
		\begin{subfigure}[t]{0.24\columnwidth}
			\centering
			\includegraphics[width=\textwidth]{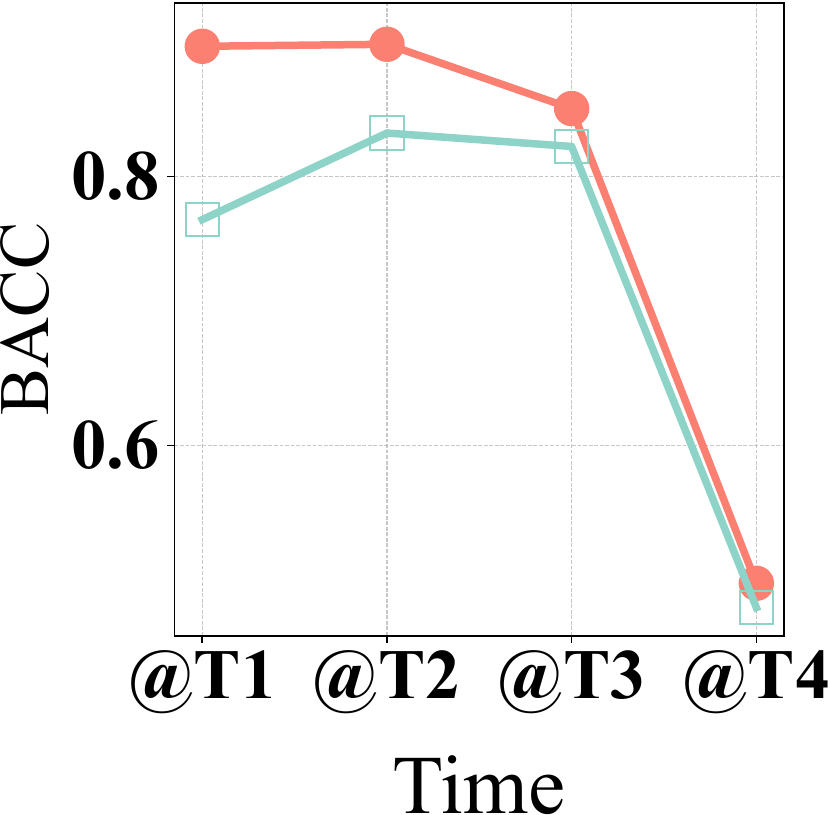}
			\caption{BACC}
			\label{f:smart:bacc_kyoto_pointwise}
		\end{subfigure}
		\hfill
		\begin{subfigure}[t]{0.24\columnwidth}
			\centering
			\includegraphics[width=\textwidth]{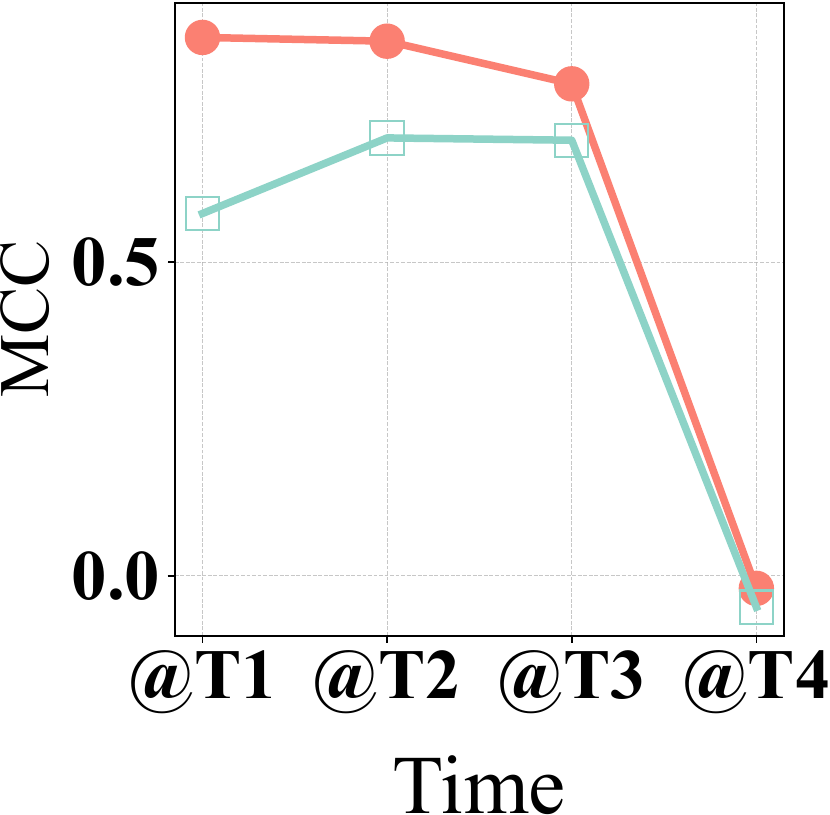}
			\caption{MCC}
			\label{f:smart:mcc_kyoto_pointwise}
		\end{subfigure}
	\end{subfigure}
	\caption{Pseudo-Labeling Comparison: Distribution-level vs. PointWise Voting mechanism on Kyoto2006+. Approaches trained on 2007 part (@T0) and evaluated after adaptation on 2011 (@T1), 2012 (@T2), 2013 (@T3), and 2014 (@T4).}
	\label{f:smart:kyotoscores_pointwise}
	\vspace{-5mm}
\end{figure}

\begin{figure}[t]
	\centering
	\includegraphics[width=0.33\textwidth]{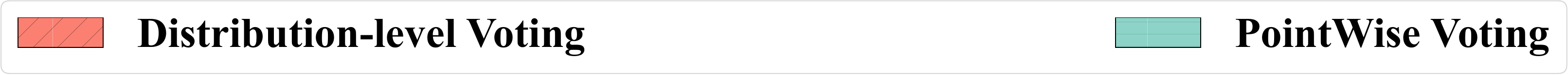}
	\begin{subfigure}[t]{\columnwidth}
		\centering
		\begin{subfigure}[t]{0.24\columnwidth}
			\centering
			\includegraphics[width=\textwidth]{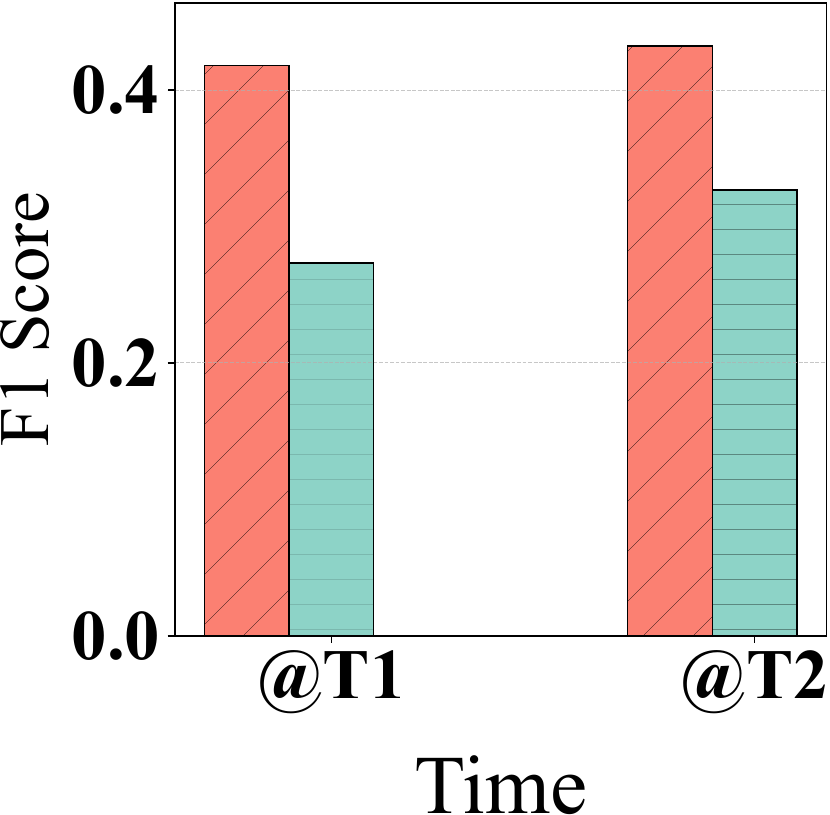}
			\caption{F1-Score}
			\label{f:smart:f1_cic_pointwise}
		\end{subfigure}
		\hfill
		\begin{subfigure}[t]{0.24\columnwidth}
			\centering
			\includegraphics[width=\textwidth]{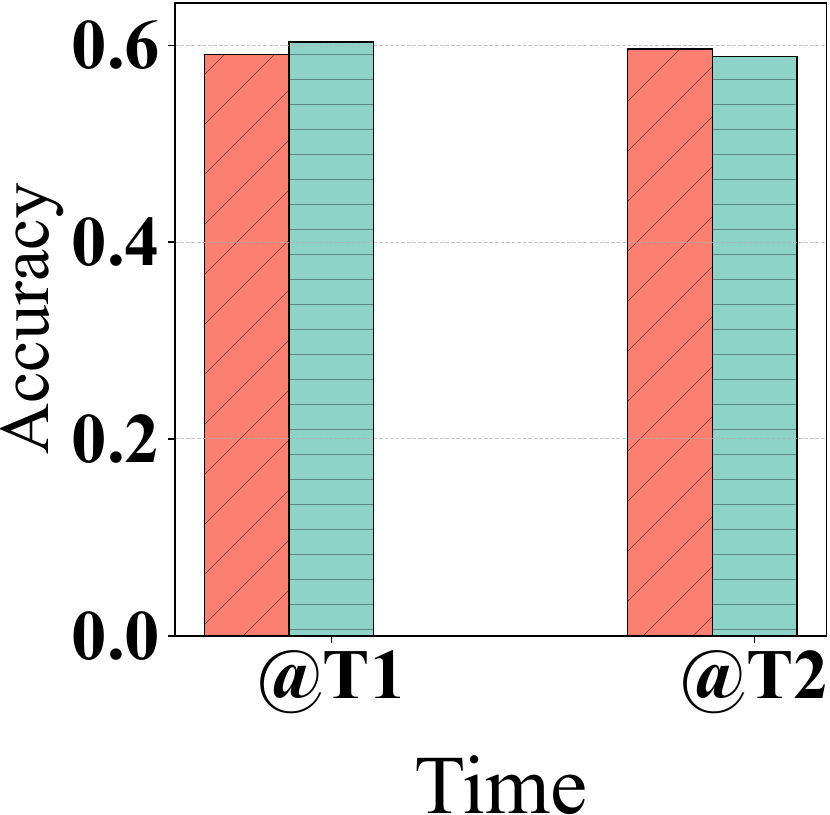}
			\caption{Accuracy}
			\label{f:smart:acc_cic_pointwise}
		\end{subfigure}
		\hfill
		\begin{subfigure}[t]{0.24\columnwidth}
			\centering
			\includegraphics[width=\textwidth]{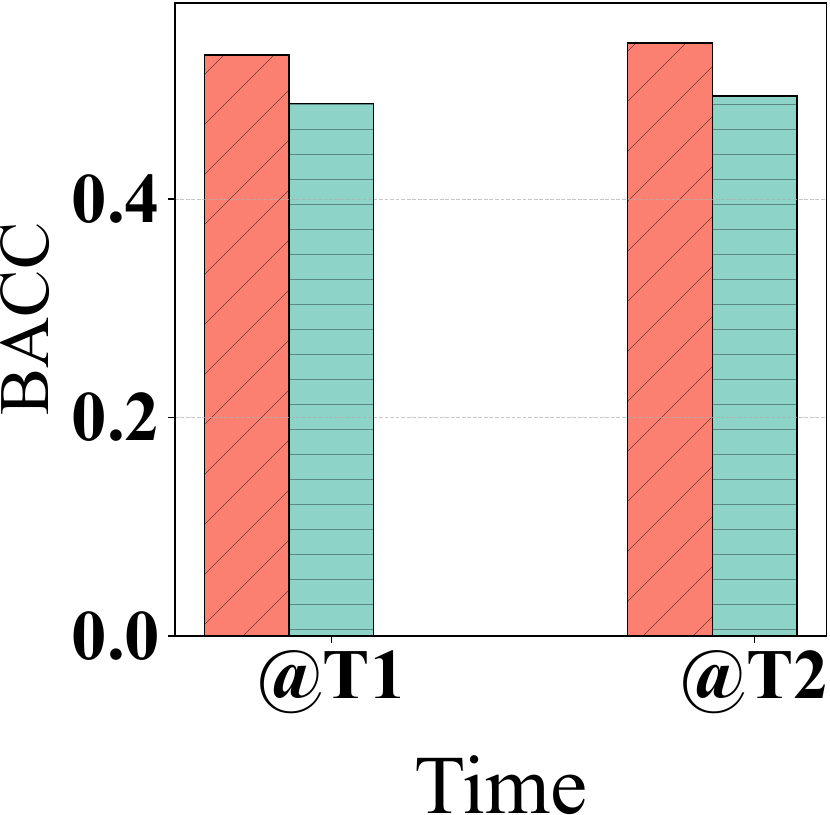}
			\caption{BACC}
			\label{f:smart:bacc_cic_pointwise}
		\end{subfigure}
		\hfill
		\begin{subfigure}[t]{0.24\columnwidth}
			\centering
			\includegraphics[width=\textwidth]{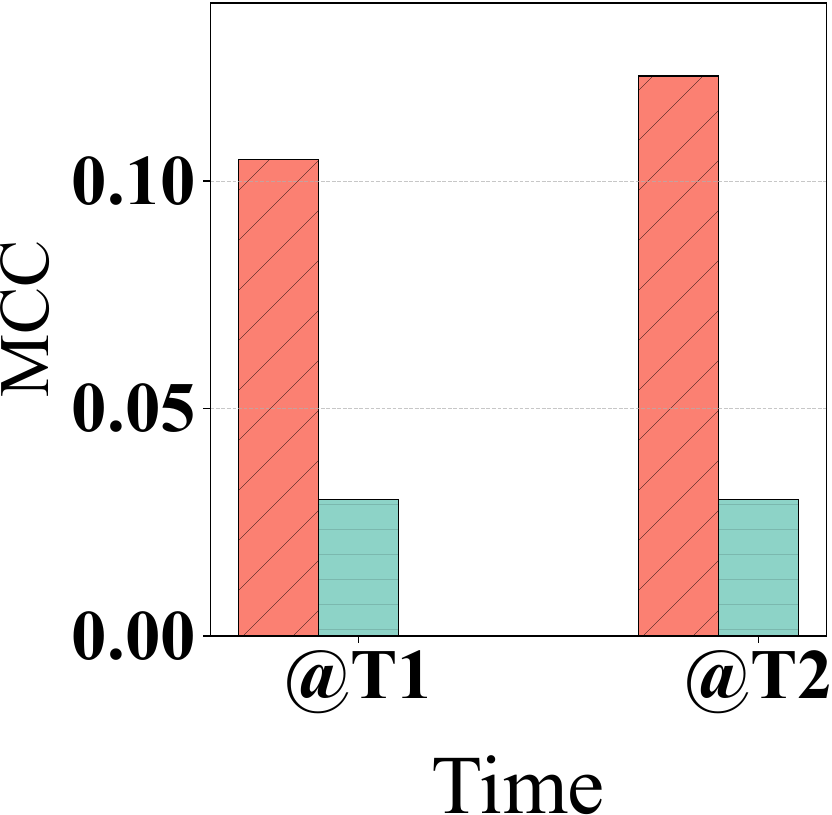}
			\caption{MCC}
			\label{f:smart:mcc_cic_pointwise}
		\end{subfigure}
	\end{subfigure}
	
	\caption{Pseudo-Labeling Comparison: Distribution-level vs. PointWise Voting mechanisim on CICIDS2017 and CICDDoS2019. Approaches trained on CICIDS2017 and evaluated before adaptation (@T1) and after adaptation (@T2) on CICDDoS2019. }
	\label{f:smart:cicscores_pointwise}
	\vspace{-5mm}
\end{figure}

For clarity, we refer to AOC-IDS's pseudo-labeling strategy as "PointWise" since it compares the distributions for each sample. Figs.~\ref{f:smart:kyotoscores_pointwise} and~\ref{f:smart:cicscores_pointwise} illustrate the comparative performance of our pseudo-labeling strategy against AOC-IDS's. As shown, \sys\ has overall superior performance. However, a slight advantage in Accuracy for AOC-IDS's pseudo-labeling is observed prior to adaptation. We attribute this to the PointWise strategy being biased towards the majority class (in this case, normal samples), which constitute 69.53\% of the CICDDoS2019 dataset. The True Negative Rate (TNR) of the PointWise approach is 78\%, which supports our claim. 

\section{Related Work}
\label{s:related}

This section reviews existing literature on adapting deep learning models to distribution shift in dynamic environments. We then briefly survey contrastive and continual learning techniques, focusing on works relevant to \sys.


\scat{Approaches to Shift Adaptation in Anomaly Detection}
A primary challenge for deployed NIDS is distribution shift, where the statistical properties of normal traffic change over time. Early deep learning approaches focused primarily on detecting this shift. Frameworks such as CADE~\cite{yang2021cade}, Transcend~\cite{jordaney2017transcend}, and TRANSCENDENT~\cite{barbero2022transcending} are effective at identifying model aging by flagging out-of-distribution samples. However, they lack automated adaptation mechanisms, often defaulting to full, manual retraining. Other methods like INSOMNIA~\cite{andresini2021insomnia} trigger retraining based on model uncertainty, but this can be an unreliable mechanism for true distributional shift. More advanced frameworks like OWAD~\cite{han2023anomaly} integrate a complete cycle of shift detection and adaptation, but its zero-positive learning basis requires clean normal training data for adaptation, which is generally difficult to obtain.


\scat{Contrastive Learning}
Several NIDS frameworks leverage contrastive learning to learn a discriminative feature space that separates normal and malicious samples~\cite{zhang2024aoc, liu2021anomalycontrastive, iot}. The primary contribution of these methods is improved initial detection accuracy by creating a tightly clustered representation of normal behavior. However, their limitation is that they confine the use of contrastive learning solely to this initial detection phase. They fail to exploit the rich, relational knowledge captured by the contrastive model during the critical adaptation phase. This leaves a significant gap, as the structural understanding of the data is not leveraged to facilitate knowledge transfer or mitigate catastrophic forgetting.

\scat{Continual Learning}
To mitigate catastrophic forgetting during adaptation, continual learning methods are often employed. Regularization-based approaches, for instance, assign importance weights to model parameters to prevent changes to critical old knowledge~\cite{du2019lifelong, han2023anomaly}. However, their primary limitation is instability, as the regularization term can dominate the main loss, leading to exploding gradients~\cite{kutalev2021stabilizing}. Memory-based methods store past data or gradients to constrain updates on new tasks~\cite{lopez2017gradient, chaudhry2018efficient}. While they prevent forgetting, their key shortcoming is strategic forgetting; they overlook the old knowledge, which leads to insufficient learning of new tasks. Some hybrid methods like SSF~\cite{zhang2025continual} combine a replay memory with knowledge distillation, but are limited by inefficient distillation, where a student merely imitates the teacher's outputs, ignoring the deeper, structural relationships within the data~\cite{li2017learninglwf, vahedifar2025no}. In summary, existing works face a dilemma between the instability of regularization and the superficial knowledge transfer of memory-based techniques.

\scat{Knowledge Distillation}
Some continual learning methods use knowledge distillation to transfer knowledge from a pre-trained teacher to a student model used in adaptation~\cite{hinton2015distillingknowledgeneuralnetwork}. In the context of NIDS, existing approaches are often limited to a naive form of distillation where the student simply imitates the teacher's final predictions~\cite{li2017learninglwf, zhang2025continual}. This approach ignores the richer, structural knowledge captured by the teacher model. While more sophisticated methods like Similarity-Preserving Knowledge Distillation (SPKD) have shown that preserving relational knowledge is more effective~\cite{tung2019similarity}, its application to the specific problem of distribution shift adaptation in NIDS remains an open challenge. Our work addresses this gap by developing a distillation strategy that preserves the pairwise similarity structure learned by a contrastive model.

\section{Conclusion}
\label{s:conc}
In this work, we presented \sys, a framework for handling normality shifts in supervised network anomaly detection. Our approach uses a novel distribution-level voting mechanism for assigning pseudo-lables and a novel similarity-preserving knowledge distillation loss to continually adapt to these shifts, effectively addressing catastrophic forgetting while learning new data patterns without costly retraining or manual labeling. Extensive experiments on three longitudinal NIDS datasets demonstrated our framework's robustness and effectiveness compared to state-of-the-art methods in this area, achieving F1-score improvements of up to $11.72\%$. 
Future directions include developing more domain-aware shift explanation methods and exploring more strategic knowledge distillation techniques to improve and enhance knowledge transfer from the old model to the new model.

\newpage
\bibliographystyle{IEEEtran}
\bibliography{main}

\end{document}